\documentclass[preprintnumbers,amsmath,amssymb,floatfix,twocolumn,showpacs,prl]{revtex4}

\usepackage{dcolumn}
\usepackage{bm}
\usepackage{graphicx}

\begin{document}


\title{Attachment of  Surface ``Fermi Arcs'' to the Bulk Fermi Surface:
 ``Fermi-Level Plumbing'' in Topological Metals}

\author{F. D. M.  Haldane}
\affiliation{Department of Physics, Princeton University,
Princeton NJ 08544-0708}

\date{January 2, 2014}
\begin{abstract}
The role of ``Fermi arc'' surface-quasiparticle states in ``topological metals'' (where some
Fermi surface sheets have non-zero Chern number) is examined.    They
act as ``Fermi-level plumbing'' conduits that transfer quasiparticles among groups of
apparently-disconnected  Fermi sheets with non-zero Chern numbers to
maintain equality of their chemical potentials, which is required by
gauge invariance.  Fermi arcs have a chiral tangential attachment to
the surface projections of sheets  of the bulk  Fermi Surface:   
the  total Chern number of each projection equals   
the net chirality of  arc-attachments to it. Information from the
Fermi arcs is needed to unambiguously determine the quantized 
part of the anomalous Hall effect that is not determined at the bulk Fermi surface.
\end{abstract}
\maketitle

The recent interest in  ``Weyl semimetals'' and the topological ``Fermi arc''
states on their surfaces\cite{ashvin} raises the question of how these
evolve as the Fermi level moves away from the  semimetal point,
and metallic behavior is restored.    If the Weyl semimetal has broken
time-reversal symmetry, it exhibits an unquantized  intrinsic
anomalous Hall effect (AHE), which
can be obtained as  a limiting case of the  general Fermi-surface formula\cite{fdmh04} for the
unquantized part of the intrinsic AHE of the metal.  The  Weyl-semimetal 
surface-state Fermi arcs survive in the metallic state as
``surface conduits'' that can adiabatically transfer quasiparticles between topologically-non-trivial sheets of the
metallic Fermi surface (FS)  that are disjoint in the bulk.    In this
Letter, I will show how they are related to, and required by,
topological features of the Fermi surface AHE formula\cite{fdmh04}.

The Karplus and  Luttinger ``intrinsic'' theory\cite{lutt} of the AHE
in ferromagnetic metals was largely ignored until it was
reinterpreted\cite{niu, nagaosa}
in modern language in terms of the geometrical (Abelian) Berry curvature of the 
spin-split Bloch bands, and is now recognized as a major component of
the AHE (in addition to ``extrinsic'' terms).     In agreement with the fundamental notion
that all transport processes occur at the Fermi level, it was
subsequently shown\cite{fdmh04}  that Karplus and Luttinger's  AHE formula could
be expressed (up to a topologically-quantized part) in terms of
Fermi-surface geometry in the Brillouin zone (BZ),
plus  the Berry-connection geometry of quasiparticle states \textit{at} 
the FS.  Note that, within Fermi-liquid theory,  the infinite-lifetime
quasiparticle states \textit{at} the $T=0$ Fermi level are the only non-topological
features of one-electron band theory that completely survive in the
presence of interactions.

The intrinsic AHE Fermi-surface formula for a  3D metal  with broken
time-reversal symmetry states that the intrinsic Hall conductivity
tensor has the form
\begin{equation}
\sigma^{ab}_{H} = \frac{\epsilon^{abc}K^H_c}{2\pi R_K},
\end{equation}
where $R_K$ = $h /e^2$  is the fundamental quantum unit of
electrical resistance, and $\bm K^H$ is a reciprocal vector with units
$[\text{length}]^{-1}$.      If a uniform magnetic flux density $\bm B$ passes
through a finite sample with volume $V$, held at fixed
electronic chemical potential $\mu$, the total electronic  charge
$Q$ obeys the Str\v{e}da  relation\cite{streda}
\begin{equation}
\lim_{V\rightarrow \infty} \frac{1}{V}\left . \frac{\partial Q}{\partial B^a}\right |_{\mu} =
\frac{K^H_a}{2\pi R_K}.
\end{equation}  
This relation requires the existence of states at the Fermi level if
$\bm K^H \ne \bm 0$ (which may be surface states).

If the  bulk material in the limit $T \rightarrow 0$ is an insulator with no Fermi surface, $\bm
K^H$ = $\bm G$ $\in \tilde \Lambda$, the Bravais lattice of reciprocal lattice
(Bragg) vectors of the bulk crystal structure.   If $\bm G$ = $\nu\bm G_0$,
where
$\nu$ is an integer, and $\bm G_0$ is primitive, this is equivalent to
a 2D  integer quantum Hall effect   $\sigma_H$ = $\nu R_K^{-1}$ in each
lattice plane indexed by $\bm G_0$\cite{kohmoto}.
From a band-structure perspective, $\bm G$ = $\bm G_{\rm QHE}$ =  $\sum_i \bm G_i$, where
$\bm G_i$  are topological invariants (an integral of the  Berry curvature)
of each disjoint group of
occupied bulk bands below the Fermi level, but it is also a
Fermi-level property of topologically-required  surface 
states  on facets of a  crystal  not normal to $\bm G_{\rm QHE}$.

If a system with no bulk states at the Fermi level has $\bm K^H$
quantized as a reciprocal lattice vector, the non-quantized part of
$\bm K^H$ must be a bulk Fermi-surface property\cite{fdmh04}.  It will be here
assumed that the (spin-split) FS is ``regular'',
\textit{i.e.},
everywhere non-degenerate, with a finite Fermi velocity,  and
described by a set of disjoint differentiable orientable
2-manifolds $\{S_i\}$ embedded in the (reduced) 3D BZ.
This is the generic case for  a ferromagnetic metal. 

A FS sheet may be parameterized by $\bm s$ $\equiv$
$(s^1,s^2)$,
 with an area 2-form ``$d^2k$" = $dA > 0$ where
$dA$ = $\frac{1}{2}(\bm n_F \cdot \partial_{\mu}\bm k_F
\times \partial_{\nu}\bm k_F) ds^{\mu}\wedge ds^{\nu}$, 
and $\bm n_F$ is the outward normal  (direction of the Fermi velocity).
 The full AHE  formula\cite{fdmh04} is
\begin{equation}
\bm K^H  = \bm G + \sum_i \int_{S_i} \frac{\bm k_F \mathcal F}{2\pi} +
\sum_{i\alpha}  \oint_{\partial S_{i\alpha}}
  \frac{\bm G_{i\alpha} \mathcal A}{2\pi} .
\label{ahe}
\end{equation}
 Here $\mathcal F$ is the Berry-curvature 2-form $\frac{1}{2}\epsilon_{abc}n_F^a\mathcal
 F^{bc}dA$ where  $\mathcal F^{ab}$ = $\nabla _k^a\mathcal A^b -
\nabla _k^b\mathcal A^a$ is the Bloch-state Berry curvature, expressed
in terms of the Berry connection $\mathcal A^a$.
 The quantity $\mathcal A$ $\equiv$ $\mathcal A^adk_a$ is
the Berry-connection 1-form on curves $\partial S_{i\alpha}$ where
sheet $S_i$ of the FS intersects the ``reduced BZ boundary'' (see below).
Note that, in the language
of differential forms, $\mathcal F$ is
the exterior derivative $d\mathcal A$.
  
It is perhaps useful to note that the Bloch-state Berry connection 
$\mathcal A_n^a(\bm k)$ = $ -i\langle \Phi_n(\bm k)
|\nabla^a_k|\Phi_n(\bm k)\rangle$ 
is
defined not just by the Bloch states  $|\Psi_n(\bm k)\rangle$ themselves, but
by $|\Phi_n(\bm k)\rangle $ = $U(-\bm k)|\Psi_n(\bm k)\rangle$, with
\begin{equation}
U(\bm k)  = \sum_{\bm R \alpha} \exp (i\bm k\cdot \bm x_{\bm
    R\alpha})|\bm R,\alpha\rangle\langle \bm R ,\alpha|,
\end{equation}
where $|\bm R,\alpha\rangle$ is an orthonormal basis of
spatially-localized orbitals in unit
cell $\bm R$, that is \textit{embedded} in  Euclidean space at $\bm
x_{\bm R\alpha}$ = $\bm R + \bm x_{\alpha}$, so the Berry curvature of
Bloch states depends not only on the details of the electronic band
structure, but also on the location of  orbitals 
within the unit cell.
This affects the the semiclassical
equations of motion\cite{sun,mar}, but  not the
topological invariants.

While the coordinate-independent formula (\ref{ahe}) 
is a simple and elegant expression, it has a number of subtleties.
First, I note that the Bloch vector $\bm k$ (and hence
the quasiparticle Fermi vector $\bm k_F$) of a charged
particle is itself ambiguous when time-reversal symmetry is broken,
as under a gauge-transformation,
 $\bm k$ $\mapsto$ $\bm k - e\bm
A/\hbar$, $\bm \nabla \times \bm A$ = 0.  In particular, the choice of a constant vector potential
$\bm A$ is compatible with Bloch states, and can only be excluded if
time-reversal symmetry is unbroken.       All physically-meaningful
(\textit{i.e.}, gauge invariant) formulas should therefore be invariant under
the mapping $\bm k_F$ $\mapsto$ $\bm k_F$ + constant.
Thus gauge invariance imposes the condition
\begin{equation}
\sum_i \int_{S_i} \mathcal F \equiv  2\pi \sum_i c_1(S_i) = 0,
\end{equation}
where the integer $c_1(S_i)$ is the ``Chern number''  of Fermi-surface
sheet $S_i$  (more technically, the first Chern class of  the mapping
between the 2-manifold $S_i$ 
and the ``$U(1)$ fiber bundle''  defined by the $\bm k_F$-dependent  quasiparticle wavefunctions inside the unit cell).

While the reciprocal-vector-valued 1-form $d\bm k_F$ is well-defined,
the reduction of $\bm k_F$ to the reduced BZ (which formally is a
3-torus with a Euclidean metric) means that if a
Fermi-surface sheet $S_i$ admits ``open orbits'' where
\begin{equation}
\oint_{\Gamma}  d\bm k_F = \bm G(\Gamma) \ne \bm 0,
\end{equation}
it is necessary for it to contain  inscribed boundary lines $\partial
S_{i\alpha}$ across which $\bm k_F$ jumps by a reciprocal vector $\bm
G_{i\alpha}$.  Then $\bm G(\Gamma)$ = $-\sum_{\alpha} \bm G_{i\alpha}$ is
canceled by the sum of jumps along the path $\Gamma$.
With the inscribed boundaries, Stokes' theorem
can be used to write $\int_S \bm k_F \mathcal F$ $\equiv$ $\int_S \bm k_F d\mathcal A$
as an integral over the interior of the intersection of $S_i$ with the
reduced BZ, plus  boundary terms.     The boundary
terms come
in matched pairs $\partial S_{i\alpha,\pm}$, which combine to give the second term in
(\ref{ahe}).  (The version of (\ref{ahe}) given in Ref.\cite{fdmh04}
sums over both ``$+$''' and ``$-$'' boundaries, so has
a prefactor $1/4\pi$ in front of the second term; since these are two sides
of the same boundary, and contribute
equal amounts, they have here been combined into a single term in (\ref{ahe}).)
The choice of the BZ boundaries on each $S_i$ is a completely-arbitrary
``gauge choice'', so  physically-meaningful results must be
invariant under a continuous change of $\partial S_{i\alpha}$: the
formula (\ref{ahe})  satisfies this requirement.

The embeddings of bulk FS sheets in the BZ also have some topological
characteristics that are independent of Berry curvature.   While 
 not directly relevant to the AHE, I list them here for completeness.
The set of open-orbits define a  Bravais lattice $\tilde \Lambda_i$
= $\{\bm G(\Gamma),\Gamma \in S_i\}$  $\subset \tilde \Lambda$. 
The Gauss-Bonnet theorem relates the integral of the
Gaussian-curvature 2-form $\kappa$ = 
$\frac{1}{2}(\bm n_F\cdot\partial_{\mu}\bm n_F \times \partial_{\nu}
\bm n_F)
ds^{\mu} \wedge ds^{\nu}$ to the genus.  Finally,
``Quasi-1D'' systems are characterized by a special primitive lattice
translation $\bm R_0$, where 
 all open orbits have $\bm
G(\Gamma)\cdot \bm R_0 = 0$,  and
some FS sheets have a 
 ``Luttinger anomaly'' (chiral anomaly)
\begin{equation}
2\pi \int_{S_i} \bm n_F dA = \Omega_{\rm BZ}\sigma_i \bm
R_0,\quad  \sigma_i = \pm 1,
\end{equation}
where  $\Omega_{\rm BZ}$ is the reciprocal-space volume of the BZ.
Such FS sheets do not enclose a definite
reciprocal-space volume, so the Luttinger theorem (relating the
geometric volume of the  Fermi surface to electron density) does not apply to them individually.
Gauge invariance requires that the total Luttinger anomaly $\sum_i
\sigma_i $ vanishes.

The gauge-invariance conditions can be strengthened in the ``ultra-clean''
limit where equilibration of the Fermi surfaces only occurs through
scattering processes with infinitesimal momentum transfer.    In that
limit, a \textit{separate} chemical potential can be 
established on  disjoint FS sheets $S_i$, which gain
\textit{separately-conserved  quasiparticle currents}.    This
requires
invariance of the AHE formula (\ref{ahe}) under a rigid
displacement of  $\bm k_F(\bm s)$ (a gauge transformation)  for the
sheet $S_i$ \textit{by itself}.   
Only  \textit{sets  of sheets  with zero total Chern number} can be
displaced together to define such  conserved currents.    

Each conserved
quasiparticle current will be associated with an independent chemical
potential,  so ``irreducible  sets'' of FS sheets associated
with a common chemical potential can be defined.      In the simplest case,
a sheet with  Chern number $+1$ is paired  with a  partner 
that has Chern number $-1$.    (The systematic classification of  the possible
structures  of ``irreducible sets'' of FS sheets is left as an open problem.)
The members of such a pair maintain a common chemical potential, so
slowly-varying external fields must be able to ``pump'' quasiparticle
charge between them without any quasiparticles being scattering
through the regions of reciprocal space that separate the disjoint
members of the pair in the BZ.           Ref.\cite{fdmh04} attributed
this to a hidden``wormhole'' connection between the two apparently separated FS sheets.

The recent work on Weyl semimetals\cite{ashvin} 
allows the mechanism for maintaining a common chemical potential on
apparently disjoint Fermi surface sheets to now be  explicitly identified with 
the Fermi arc surface states.  
In Weyl semimetals, 
the Fermi surface collapses to a set of discrete points in the
BZ, which are band-touching degeneracies.  The  Weyl points at $\bm k^0_i$  are monopole
sources of ``Berry flux'' which absorb a multiple of $2\pi$ of Berry
curvature flux from one band, and emit it into the other.   If the Fermi
level  is slightly shifted above or below the Weyl points, regular Fermi
surfaces surrounding each Weyl point emerge, with Chern numbers
$c_1(S_i) \equiv c_i$.
The limit of $\bm K^H$ as the Fermi level passes through  the Weyl
points is just
\begin{equation}
\bm K^H  = \bm G  + \sum_i c_i \bm k^0_i .
\end{equation}

A simple ``toy model'' for the Fermi arcs is provided by a 1D
model often used  to model  a ``quantum pump'':
\begin{equation}
H = \sum _{n=1}^{\infty} (-1)^nV c^{\dagger}_nc_n + 
\sum_{n,\pm}(t_{\pm}c^{\dagger}_{2n\pm 1}c_{2n} +\text{H.c.}).
\label{model}
\end{equation}
The two bulk bands are $\varepsilon_{\pm}(k) = \pm \surd ( V^2 +
t_+^2 + t_-^2 + 2t_+t_-\cos k)$.    I will take $t_+ > 0$;
the band gap then closes at $k$ = $\pi$ for $t_+$ = $t_-$ and $V$ = 0.
In general, the gap is $|E|$  $<$ $\surd \left ( V^2 + (t_+-t_-)^2\right)$, and 
for $|t_-| <t_+$, there is an edge-state in the gap with $\Psi_{2n}$ = 0, $\Psi_{2n+1}$ 
= $(-t_-/t_+)^n\Psi_1$, and $E$ = $V$.   By making $V$, $t_-$ and
$t_+$ functions of the surface Bloch vector $\bm
k$ = $(k_x,k_y)$, this
can   model  Weyl points and their Fermi
arcs\cite{hosur}.

\begin{figure}
\includegraphics[width=8.5cm]{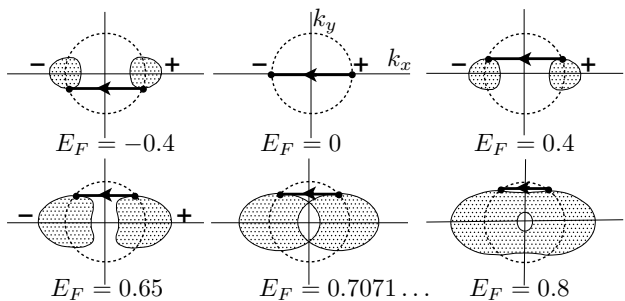}
\caption{
Surface-state Fermi arcs (bold lines) (with directions $\bm n\times
\bm v_F$) and their tangential attachment to
the projections (shaded) of the bulk Fermi surface  into the surface Brillouin
zone $\bm k$ = $(k_x,k_y)$, for the ``toy model'' (\ref{model}) with
$(V,t_-,t_+)$ = $(k_y,|\bm k|, 1)$.
An incomplete band of chiral surface
states with  $E(\bm k)$ =
$k_y$ exists in the region $|\bm k|$ = $\exp -\kappa/2$  $< 1$ (bounded
by the dotted circle); $\kappa^{-1}$ is the decay length into the bulk.
Fermi arcs exist for  $|E_F|  <1$; the bulk Fermi surface
splits into two  sheets
  with Chern numbers $\pm 1$ for $|E_F| < 1/\surd 2$, and collapses to
  two Weyl points as $E_F \rightarrow 0$. 
}
\label{fig1}
\end{figure}

As an example, Fig.(\ref{fig1}) shows results for model parameters $(V,t_-,t_+) $ = $(k_y, |\bm k|,1)$
The surface state has dispersion $E(\bm k)$ = $k_y$, and
exists for $|\bm k|< 1$, and its decay length into the bulk
diverges as $|\bm k| \rightarrow 1$.   For Fermi energy
$E_F$ = 0, the model is a Weyl semimetal with Weyl points at $\bm
k$ = $(\pm 1,0)$.   For $E_F \ne 0$, the  boundaries of the
region  of the projected bulk FS are
given by
\begin{equation}
(k_y)^2 + (|\bm k| - 1)^2 = ( E_F)^2,
\end{equation}
and shown in Fig.(\ref{fig1}) for selected values of $E_F$.   For
$| E_F| < 1$ a Fermi arc of 2D quasiparticles connects two points
on the surface of the projected FS, and emanates tangentially from it.

The ``toy model'' highlights a number of features.   First, its surface
states are an ``incomplete band'', as they exist in only a limited
region of  $k$-space, and would not cover the full two-dimensional
surface BZ (2DBZ) in a realistic
model that was periodic parallel to a crystal facet.
Instead, the surface band  terminates on a $k$-space boundary at which
the decay-length into the bulk diverges.   Close to the boundary, the
surface state is extremely weakly-bound, and its properties approach
those of the bulk electronic band from which it evolves at the
termination point.     In particular, its group velocity tangent to
the surface will approach that of the bulk band edge at the
termination point from which it evolves.   

The end points of Fermi arcs are at the
intersection of the projected bulk FS with   the
termination line where an incomplete surface band  leaks into the
bulk,
 and the attachment
is  thus generically
tangential.  In the 2DBZ, the Fermi-vector 1-form $d\bm k_F$ can be
given a standard direction so that $\bm n\times  \bm v_F \cdot d\bm
k_F$ $ > 0$. where $\bm n$ is the outward normal of the facet, and
$\bm v_F$ is the surface quasiparticle Fermi velocity tangent to the facet.
 At the attachment point $\bm k_ i$, there are two possible tangential
 directions for the arc to leave the attachment point: $\xi_i d\bm k_F$
 where $\xi_i$ = $\pm 1$.   By inspection, there a sum rule
\begin{equation}
\sum_{i\in P_{\rm FS}} \xi_i =  \sum_{i\in P_{\rm FS}} c_i(S_i)
\label{sumrule}
\end{equation}
where the LHS is the the sum of attachment chiralities to a given
region $P_{\rm FS}$ of projected FS in the facet BZ, and the RHS is the
sum of Chern numbers of the bulk FS sheets contributing to the projection.
If this is non-zero,  a net number of  directed Fermi arcs
flow towards or away from   $P_{\rm FS}$, and must terminate on other projections with
compensating Chern numbers.

There is no need for a projected FS  to have a
topologically-non-trivial Chern number for it to be attached to a
Fermi arc, as the ``toy model'' shows for $1/\surd 2 < |E_F| < 1$.
Arcs that detach and re-attach to the same FS projection
do not affect the LHS of (\ref{sumrule}).

The arcs define a 1D \textit{open} manifold of surface  quasiparticle
states embedded in the facet 2DBZ.     \textit{Closed} surface  quasi-particle manifolds
$C_i$ (unconnected to the Bulk FS projections) can also be present: in
particular, the chiral edge states 
deriving from completely occupied bands, with no bulk connection to the
Fermi surface, that exhibit a 3D integer quantum Hall effect. 
These are
associated with a net 2D Luttinger anomaly
\begin{equation}
\bm R \cdot \sum_i \oint_{C_i} d\bm k_F = \bm R\cdot \bm G_{\rm QHE},
\end{equation}
where $\bm R$ is any lattice translation parallel to the facet.

The only residual reciprocal lattice vector ambiguity in (\ref
{ahe})
 is associated with  irreducible groups of disjoint Fermi
sheets individually carrying a non-zero Chern number.     The AHE
formula is essentially a dipole moment of FS Berry curvature in the
BZ, which is ambiguous if  sheets carrying quantized Chern number ``charges'' are shifted
relative to the others by a reciprocal lattice vector.    To define
the total ``dipole moment'' of the group, they must be placed together
in the interior of some choice of reduced BZ.

The simplest ``correct'' choice is to choose the reduced BZ so that when
projected into the 2DBZ of a facet, no Fermi arcs cross its boundaries.
However, a general formula should not
make special ``gauge choices'':  the relation between
this and an arbitrary choice of reduced BZ is found by  projecting the
reduced BZ boundaries into the facet and summing over the intersections of
the directed Fermi arcs with the projected BZ boundary, weighted by the jump
$\Delta \bm k_F$ = $\bm G_i$ in the 2DBZ direction $d\bm k_F$.  Thus the value of $\bm G$ in (\ref{ahe})
can be \textit{unambiguously} determined from the inspection of the closed
Fermi curves and open Fermi arcs in the 2DBZ of the facets.    

The 2D
surface Hall conductivity, also given in Ref.\cite{fdmh04},  will be the next-to-leading term in a
formula for a large (but finite) crystal:
\begin{equation}
\left . \frac{\partial Q}{\partial B^a} \right |_{\mu}
= \frac{1}{2\pi R_K} \left ( V K_a + \sum_{\alpha} \theta_{\alpha}
  A_{\alpha} n^{\alpha}_a\right ) .
\label{extended}
\end{equation}
Here $A_{\alpha}$ is the surface area of facet $\alpha$ and $\bm
n^{\alpha}$ is its outward unit normal (these always obey $\sum_{\alpha}
A_{\alpha}\bm n^{\alpha}$ = 0).
The 2D AHE formula\cite{fdmh04} states that
\begin{equation}
\theta_{\alpha}
 = 2\pi \nu_{\alpha} + \sum_i \oint_{C_{i\alpha}} \mathcal A
\end{equation}
where $\mathcal A$ = $\mathcal A^adk_{Fa}$ is the Berry connection
1-form on the closed directed  Fermi curve $C_{i\alpha}$ in the facet
2DBZ. The closed  Fermi curves in the
reduced 2DBZ determine the unquantized value of $\theta_{\alpha} \mod 2\pi $.  The quantities
$\nu_{\alpha}$ are integers, and are determined  by integer quantum Hall 
edge states (free electron chiral Luttinger liquids with an integer
chiral anomaly) that may be
present on the edges between adjacent facets.

Closed Fermi curves in the reduced 2DBZ of a facet  are disconnected
from the  bulk Fermi surfaces  and can thus (on a clean
surface) support independent
chemical potentials,  justifying the additional surface term they
contribute in (\ref{extended}).
In contrast,
the 2D AHE formula of Ref.\cite{fdmh04} has no obvious place for a
contribution to the subleading surface-AHE from open Fermi arcs, since only  
closed 1-manifolds can have a gauge-invariant Berry phase factor $\exp
i\oint \mathcal A$.        In addition, the Fermi arcs do not have
chemical potentials independent of the bulk FS sheets they attach to,
so
it seems consistent that, as well as not contributing to the
non-quantized part of the bulk
AHE (in contradiction to recent claims\cite{cbb}),
 they will also not contribute any
independent extra non-quantized terms to the
sub-leading (facet) terms of (\ref{extended}).   

As a final example, consider a system which has a trivial
insulator bulk with $\bm K$ = 0, and where all facets are 2D Chern
insulators.   In this cases the only states at the Fermi level are
chiral 1D Fermi liquids on the edges of the facets, which form a
network of 1D edges, each of which has a directed integer chiral anomaly 
$\nu$.    The edges are joined at the crystal vertices, and the
net outgoing chiral anomaly on edges leaving a vertex must vanish.
This means that the formula (\ref{extended}) has the correct form, as
it can then be decomposed into a sum of facet terms, each of which
contributes an quantized integer QHE term   $\nu_{\alpha}
A_{\alpha}n^{\alpha}_a/  R_K$
to (\ref{extended}).  The absolute value of $\nu_{\alpha}$ is determined
on each edge by the chiral anomaly, the number of
``right-moving'' minus the number of ``left moving'' Fermi points, 
 which can each have its own chemical potential.

The preceding discussion  assumes that the Fermi surfaces in
the 3D BZ, as well as the closed Fermi curves and open Fermi arcs in the
facet 2DBZ, are described by Fermi-liquid theory.    In the
final example, where the only gapless excitations derive from integer quantum Hall edge
states on the  crystal edges between facets, there is a natural interaction-based
generalization to fractional quantum Hall states, with fractional
Chern insulator facets.    An interesting open question
remains:
can the gapless 2D or 3D
Fermi-liquid states, which provide non-quantized
geometric parts
of the formula (\ref{extended}), also have non-trivial
generalizations
in strongly-interacting systems?

Acknowledgements.— This work was supported by the NSF MRSEC
DMR-0819860 at the Princeton Center for Complex Materials.
This work was also partially supported by a grant from the Simons Foundation
(\#267510 to Frederick Haldane)  for Sabbatical Leave support.

\end{document}